# Emergence of partially disordered antiferromagnetism and isothermal magnetization plateau due to geometrical frustration in a metallic compound, $Er_2RhSi_3$

Kartik K Iyer,[1,2] Kalobaran Maiti,[1] Sudhindra Rayaprol,[3] Ram Kumar,[4]
S. Mattepanavar,[5] S. Dodamani,[2] and E.V. Sampathkumaran[6]

[1]*Tata Institute of Fundamental Research, Homi Bhabha Road, Colaba, Mumbai – 400005, India*
[2]*KLE Society's Dr. Prabhakar Kore Basic Science Research Centre, KLE Academy of Higher Education and Research, Belagavi- 590010, India*
[3]*UGC-DAE-Consortium for Scientific Research -Mumbai Centre, BARC Campus, Trombay, Mumbai – 400085, India*
[4]*Maryland Quantum Materials Center, University of Maryland, College Park, MD20742, USA.*

[5]*KLE Society's, Basavaprabhu Kore Arts, Science & Commerce College Chikodi-591201, India.*
[6]*Homi Bhabha Centre for Science Education, TIFR, V. N. Purav Marg, Mankhurd, Mumbai – 400088, India*

'Partially disordered antiferro (PDA) magnetism' (in which one of the three magnetic ions in a triangular network remains magnetically disordered), has been known commonly among geometrically frustrated insulating materials. The 1/3-plateau in isothermal magnetization ($M$) of such materials has been of great theoretical interest. Here, we report these properties in a $AlB_2$-structure derived *metallic* material, $Er_2RhSi_3$, in which Er sublattice has triangular networks. The presence of a well-defined λ-anomaly in the temperature ($T$) dependence of heat capacity and its magnetic-field ($H$) dependence, and the loss of spin-disorder contribution in electrical resistivity (ρ) confirm antiferromagnetic order below ($T_N=$) 5 K. On the other hand, the separation of zero-field-cooled and field-cooled dc magnetic susceptibility (χ) curves, decay of isothermal remnant magnetization and the frequency dependence of real and imaginary components of ac χ suggest the onset of spin-glass freezing concomitant with the antiferromagnetic order. In addition, interestingly, we observe 1/3-plateau in $M(H)$ below 20 kOe for $T < T_N$. The change in ρ as a function of $H$ at a given temperature well below $T_N$ is also revealing, with this compound exhibiting a plateau below 20 kOe, with complexities at higher fields. Therefore, this compound serves as a prototype for theoretical understanding of transport behavior across 1/3 plateau due to PDA magnetism in a metal without any interference from the 4f delocalization phenomena.

The phenomenon of highly frustrated magnetism due to the geometrical arrangement of the magnetic ions, labelled 'geometrically frustrated (GF) magnetism', is one of the modern topics of research in condensed matter physics, as such a frustration has been known to lead to a variety of interesting magnetic states. [1-9]. This article essentially focusses on two of these aspects, viz., Partially Disordered Antiferromagnetism (PDA) and magnetization plateaus. The concept of PDA magnetism was originally invoked to describe a situation for materials containing triangular magnetic framework, in case the magnetic ions at two vertices are antiferromagnetically coupled, while the third one is left random due to geometrical frustration [1, 2]. This kind of magnetism has been of interest due to the fact that all the magnetic ions are crystallographically equivalent, yet exhibiting different magnetic behavior. Very early examples for PDA magnetism are $CsCoCl_3$, and $CsCoBr_3$ [1, 2], and during last two decades $Ca_3CoRhO_6$ [3-5] and $Ca_3Co_2O_6$ [6-8] have been attracting attention. With respect to the 'magnetization plateaus', this is in general a subject of considerable theoretical and experimental investigations among Shastry-Sutherand, square, kagome and triangular insulating systems, as multiple plateaus can arise not only due to GF, but also due to various other magnetic interaction frustrations [see, for instance, Refs. 9-12]. Quantum effects have been shown to play a major role. The PDA magnetic systems arising from GF in triangular magnetic lattices, are often characterized, in particular, by a plateau at the 1/3 of saturation magnetization ($M$). There is a flurry of activities in recent years for this 1/3 plateau, e.g., $Ba_3NiSb_2O_9$, $Ba_3NiNb_2O_9$ [13, 14], Volorthite [15], $Cs_2CoBr_4$ [16], $CoGeO_3$ [17], and $Cu_3Bi(TeO_3)_2O_2Cl$ [18]. The readers may also see the seminal theoretical work by Chubukov and Golosov [19]. Such plateaus occur irrespective of the spin-value of the magnetic ion. Besides many such insulators, a semiconducting triangular system $Eu_3InAs_3$ has been reported very recently [see, for instance, 11, 12, 17-34 and articles cited therein].

Clearly, the identification of materials exhibiting such quantized steps, in particular 1/3-plateau, is an important direction of research in the field of GF magnetism. Needless to emphasize that any kind of PDA magnetic structure even in other GF families is of great interest in condensed matter physics, e.g., $Gd_2Ti_2O_7$ [35]. However, such reports in a metallic environment are scarce and sporadic, e.g., $UNi_4B$ [36], CePdAl [37], $Tb_3Ru_4Al_{12}$ [38], and very recently $EuRh_2Al_8$ [39]. The compound $TbRh_6Ge_6$ [40] has been recently reported to show a 1/9-plateau, in addition to a 1/3-plateau. The question therefore arises whether the Rudermann-Kittel-Kasuya-Yosida (RKKY) indirect exchange interaction usually mediating magnetic ordering in metals, in particular among rare-earths (R), is not generally favorable to cause these features of geometric frustration. It is therefore of great interest to search for materials - without interference from other exotic phenomena due to f-electron delocalization (as in CePdAl [37]) - exhibiting features characteristic of PDA magnetism and 1/3-plateau to provide relatively simple examples to enable the theorists to work further in this direction. In this article, we present an evidence for such magnetic characteristics in $Er_2RhSi_3$, derived from $AlB_2$-derived hexagonal structure [41].

A good number of ternary rare-earth compounds of the type $R_2(TM)X_3$ (where TM is a transition metal ion, X= Si, Ge) have been derived from $AlB_2$ hexagonal crystal structure. The sites for TM and X are 4(f) and 12(i), respectively forming TM-X layers. The layers of TM-X form a honeycomb network and the layers of triangular network of R ions (Fig. 1a) alternate with TM-X along c-direction. In the event of the ordered replacement of the boron site by TM and X, two types of R, viz., 2(b) and 6(h), can be visualized depending on whether the nearest hexagons in the adjacent layers contain ordered TM/Si or X atoms. As a result of this difference in the chemical surrounding, the interatomic distances undergo subtle differences, as discussed in Ref. 42 for $Er_2RhSi_3$ (as shown in Fig. 1a for Er-Er distances). This is presumably responsible for doubling of the unit-cell parameters with respect to that expected for the disordered distribution of TM and X [in which case the chemical formula can be written as $R(TM)_{0.5}X_{1.5}$]. For crystallographic details, the readers may see Ref. 41 and Supplemental Material [42]. Among these, many novel magnetic and transport anomalies have been reported [see, articles cited in 43-56] for the past three decades for several members of $R_2PdSi_3$ family [see, articles cited in 43-56]. The most notable one is $Gd_2PdSi_3$, the transport anomalies of which include Kondo-like electrical resistivity ($\rho$) [46]; it is intriguing that Hall anomaly typical of 'Topological Hall Effect' was reported two decades ago much before this concept was recognized in metals as brought out in Ref. 49. $Gd_2PdSi_3$ is the first centrosymmetric case to exhibit magnetic skyrmion behavior [48, 50-52], strongly modulated by Pd/Si superlattice [51]. Therefore, it is important to investigate Rh analogues. Other than some initial studies long ago [42, 57, 58], and in particular on Ce case in depth [59-67], Gd, Tb and Dy members [68] are special as these exhibit magnetic and transport properties comparable to those of $Gd_2PdSi_3$. The properties of the isomorphous Er compound [69] presented in this paper bring out uniqueness in the context of PDA magnetism.

Polycrystalline sample was prepared by melting together stoichiometric amounts of constituent elements in an arc furnace in an atmosphere of argon under partial pressure, followed by annealing at 1073 K for about a week in an evacuated sealed quartz tube. X-ray diffraction pattern (recorded with Cu $K_\alpha$ radiation) was analyzed by Rietveld refinement method, which confirmed single phase nature of molten ingot (see Supplementary Material [42]). The pattern was found to be in excellent agreement with that of Ref. 41, including the appearance of superstructure lines (Fig. 1b) establishing doubling of unit-cell parameters ($a$= ~8.1043 and $c$= ~7.7518 Å). Details of temperature ($T$) and magnetic-field ($H$) dependencies of ac and dc $M$, heat-capacity ($C$), and electrical resistivity ($\rho$) measurements can be found in Ref. 42.

In the inset of Fig.1c, we show the $T$-dependence of inverse susceptibility ($\chi$) obtained in a field of 5 kOe. The plot is found to be linear over a wide $T$-range well above 10 K. The effective moment ($\mu_{eff}$) obtained from the slope of the plot (~9.7 $\mu_B$) is almost the same as the theoretical value of 9.7$\mu_B$ for trivalent Er ions. The paramagnetic Curie temperature is found to be ~1.4 K; the positive sign is indicative of ferromagnetic correlations between Er ions, which is different from other rare-earth cases in the same family [68]. We noted that, as the temperature is lowered, there is an upturn of $\chi$ around 5 K beyond the value expected from the high temperature Curie-Weiss behavior followed by a peak, as though antiferromagnetic ordering sets in, as shown in Fig. 1c for 10 Oe measuring field. The zero-field-cooled (ZFC) and field-cooled (FC) curves for this low-field tend to deviate from each other at the onset of magnetic order itself, with the FC curve continuing to rise down to 2 K, typical of spin-glasses in many concentrated magnetic materials [70-72]. For another interesting $\chi$ behaviour in the vicinity of the $T$-region 2-20 K at low-fields (20, 50 and 100 Oe), the readers may see Supplementary Material [42].

In order to explore further the origin of the above features, we have measured ac $\chi$ in the low temperature region with frequencies 1.3, 13, 133 and 1333 Hz and the results obtained as a function of $T$ in the vicinity of the magnetic transition are shown in Fig. 2. The plot is restricted to the vicinity of the magnetic transition, as the curves are featureless at higher temperatures. It is apparent from this figure that not only the real part ($\chi'$), but also the imaginary ($\chi''$) part, exhibits prominent peaks in magnitude - with the sharp upturn occurring at 5 K - as expected for spin-glasses [73]. There is a weak frequency ($\upsilon$) dependence of the peak temperature, say, in $\chi'$, that is, 0.2 K for a variation of $\upsilon$ from 1.3 to 133 Hz. [The peak values also undergo a decrease with increasing $\upsilon$, with significant suppression for 1333 Hz which is curious]. The peaks vanish for a small application of a dc magnetic field (say, 5 kOe), as revealed by the flatness of the plot in the figure. The results establish that this compound undergoes spin-glass freezing. The most notable observation, as inferred from the peak temperature in $\chi'$, is that the spin freezing occurs exactly at the onset of magnetic ordering. In order to render further support to the existence spin-glass freezing, we have measured isothermal remnant magnetization ($M_{IRM}$) as a function of time ($t$) at 2 K. This curve, shown in the inset of Fig. 2, was obtained as follows: After zero-field-cooling of the specimen at 2 K, the specimen was left in a magnetic-field of 5 kOe for 5 minutes; immediately after switching off the field, $M_{IRM}$ was measured as a function of $t$ for about an hour. From the inset of Fig. 2, it is clear that there is a slow decay of $M_{IRM}$, varying essentially logarithmically with $t$ (barring slower decay for initial few seconds) consistent with what is expected for spin-glasses.

Having established the onset of spin-glass freezing at 5 K, a support for long-range magnetic order pointing to a well-defined magnetic structure comes from the $C(T)$ data, apart from the appearance of magnetic Bragg peaks in the neutron diffraction data [57]. We show the $C(T)$ data in the $T$-region of interest only (<< 10 K) in Fig. 3a, as there is no worthwhile feature at higher temperatures, as measured up to 150 K. It is evident from Fig. 3a that the zero-field data exhibit a strong λ-anomaly with a sharp upturn at 5 K. This is a characteristic feature of long range magnetic ordering arising from a well-defined magnetic structure. Spin-glass freezing alone would have resulted in smearing of the feature at the onset of freezing. Therefore, viewing together with the features in the ac and dc magnetization presented above, this establishes that spin-glass freezing and a well-defined magnetic structure set in essentially at the same temperature, namely, at 5 K. With respect to the behavior in the presence of external fields, there is a gradual smearing and broadening of the peak and the peak moves towards a lower temperature, for example, for 30 and 50 kOe to 4.5 and 3.8 K respectively. This establishes that the strong λ-anomaly arises from antiferromagnetism. In short, these results establish that PDA magnetism occurs at the magnetic ordering temperature of ($T_N$=) 5 K. Finally, we have also derived magnetic contribution, $C_m$, to $C(T)$ by measuring $C(T)$ of La analogue as in Ref. 68, which is shown in Fig. 3b. The magnetic entropy, $S_m$, (Fig. 3b, inset) obtained by integrating $C_m/T$ versus $T$ is ~8 J/mol K at $T_N$. If one assumes that the magnetic ordering arises from the crystal-field-split doublet ground state (as $Er^{3+}$ is a Kramers ion), the minimum expected value for the magnetic entropy at $T_N$ should be 2Rln2= 11.52 J/mol K. Therefore, the observed lower value supports that a significant fraction of Er ions are magnetically disordered. It may be remarked that the theoretical value of magnetic entropy for full degeneracy of 4f orbital of $Er^{3+}$ (for which total orbital angular momentum is $J$= 15/2) is equal to (2Rln16=) 46.1 J/mol K, which is far above the observed value at $T_N$. Clearly, crystal-field splitting of 4f orbital is present, which is also supported by a similar lower value of magnetic moment determined by neutron diffraction [57].

We have derived isothermal entropy change, defined as $\Delta S = S(H) - S(0)$, from the area under the curves of $C/T$ versus $T$, measured at different fields, and the results obtained are shown in Fig. 3c. The curves fall in the negative quadrant with a negative peak, typical of a dominant ferromagnetic component in such fields [74]. This demonstrates field-induced changes in the magnetic structure in the magnetically ordered state. The peak values are reasonably large, say, for 0→50 kOe field-variation, with the curve spreading over a wide $T$-range above $T_N$. Surprisingly, $\Delta S$ changes sign sharply at the loss of magnetic order above 5 K for 10 kOe, and the positive sign above $T_N$ implies possible field-induced magnetic fluctuations in such low fields. The sign becomes negative in the paramagnetic state for $H$ >10 kOe as expected, but the large magnitude over a wide $T$-range may be due to the antiferromagnetic clusters giving rise to an effective ferromagnetic alignment. The results overall imply that this compound may be an example for interesting magnetic precursor effects [46, 47, 74, 75] in the paramagnetic state, which is also theoretically addressed in recent times [47, 77, 78].

We present isothermal magnetization behavior for 1.8, 4 and 6 K in Fig. 4a-c to bring out 1/3-plateau. There is a sharp rise at 1.8 K for the application of an initial small $H$, and there is a plateau immediately thereafter till 20 kOe. The sharpness of these features is a bit smeared in the 4 K plot. There is an upturn near about 20 kOe after the plateau and the variation is weak beyond about 30 kOe. The plot of $M(H)$ tends to flatten, varying weakly beyond 60 kOe till the measured field of 120 kOe (Fig. 4a, inset), as though there is a tendency to saturate. All these $M(H)$ curves are found to be non-hysteretic. A linear extrapolation of the high-field curve to zero field yields a value of about 6.5 $\mu_B$/Er. This value is far below the saturation value expected for fully degenerate trivalent Er ions (9 $\mu_B$ per Er) and so the reduced value can be attributed to a decrease in the magnetic moment and possible strong anisotropy resulting from crystal-field effects. The most intriguing observation relevant to the aim of this article is that the value of the similarly extrapolated magnetic moment in the plateau region is ~2.15$\mu_B$/Er, which is essentially one-third of the extrapolated value from the high-field data (obtained above). A simple-minded picture for the origin of 1/3 magnetization plateau, advanced for a triangular lattice, $Ca_3Co_2O_6$ [Ref. 5], is as follows: The magnetic ion at one of the three vortices of the triangle is magnetically disordered as evidenced by spin-glass features discussed above, while the other two are coupled antiparallelly in zero field in the virgin state; in the plateau region, the magnetic moment of the 'disordered' magnetic ion gets oriented along the field leaving the other two antiparallelly aligned to each other leading to a ferrimagnetic state. It is not straightforward to understand why such an intermediate state is stable against perturbation by a magnetic field, in our case, up to 20 kOe; presumably, the antiparallel interaction is too strong to flip the moments of Er with this external field. Beyond 30 kOe, all the three sites tend to align ferromagnetically resulting in thrice of magnetic moment of the plateau region. For the sake of a reader, a schematic representation of this scenario is shown for a triangle in Fig. 4. Thus, the 1/3 isothermal magnetization plateau could be consistently interpreted with the idea of 'Partially Disordered Antiferromagnetism'. Note that the plateau vanishes as soon as the long-range magnetic order regime is crossed, as seen in the $M(H)$ curve for 6 K.

Previous neutron diffraction studies [57] suggest that there is no crystal structure change in the magnetically ordered state. In that report, both the Er ions are assumed to carry an equal magnetic moment of 5.9 $\mu_B$ per Er, which is below the value of 9$\mu_B$ for the fully 4f-degenerate trivalent Er ion, suggesting the presence of crystal-field splitting. We would like to state that it is difficult to reconcile PDA magnetic structure proposed here with the conclusions obtained from these neutron diffraction studies. It is not uncommon, particularly within this ternary family, that reinvestigation by modern facilities and/or better quality samples have yielded results contradicting previous magnetic structure proposals, for example, for $Tb_2PdSi_3$ [79, 80] and for $Nd_2PdSi_3$ [44, 79]. Besides, a PDA system would be characterized by a broad background below the sharp magnetic peaks, suggestive of quasielastic peaks arising from spin-glass component, as in $Ca_3CoRhO_6$ [81, 82]. A careful look at the background below the intense peaks in the reported neutron diffraction pattern [57] indeed provides an indication for the same. In view of these, we call for detailed neutron diffraction studies to search for some kind of PDA magnetic structure, before attaching a significance to the apparent discrepancy with the present conclusion.

We now demonstrate the behavior of magnetoresistance (in the transverse geometry) across 1/3 plateau regime in this metallic system, serving as a food for thought for further theoretical work to understand transport anomalies in the 1/3 plateau regime, as such an opportunity is not provided by insulating PDA systems. The ρ(*T*) plot (see Fig. 3c, inset) shows distinct evidence for the loss of spin-disorder contribution to ρ at $T_N$. Figures 5a-e show the behavior of isothermal magnetoresistance, MR, at various temperatures across $T_N$. Here MR is defined as [ρ(*H*) – ρ(0)]/ρ(0). The curves are essentially nonhysteretic. Looking at the MR curve at 2.2 K (see Fig. 5a), the magnitude is negligibly small below 20 kOe, but a negative sign emerges after a small change in applied *H* (possibly due to grain boundaries or spin-glass component); it is notable that MR otherwise is almost flat in the 1/3 magnetization plateau region, as though magnetism is topologically protected, apparently avoiding even conventional magnetic and non-magnetic contributions for scattering process. After this plateau region is crossed, there is a distinct upturn to positive quadrant, followed by a weak drop around 30 kOe possibly due to ferromagnetic alignment contribution. Further increase of *H*, that is, in the region where there is a tendency for magnetization to saturate, MR surprisingly keeps increasing remaining in the positive quadrant. Such a linear MR variation in the essentially ferromagnetically aligned state is puzzling (as ferromagnets are usually characterized by negative MR with a non-linear dependence on *H*). It is not clear whether the classical contribution due to conduction electrons dominates ferromagnetic contribution at higher fields, leading to positive MR. At 4 K, the features are similar, except that the initial drop is more prominent. Just above $T_N$, say at 10 K, there is a competition between well-known paramagnetic contribution (negative sign with $H^2$-dependence) and positive classical contribution, leading to a minimum at the intermediate field range. With increasing *T*, since paramagnetic contribution should get gradually weakened, the minimum gradually vanishes, as shown for 10, 15 and 25 K (Fig. 5 b-d).

In conclusion, we report the results of our bulk measurements on $Er_2RhSi_3$ containing essentially localized 4f electrons in a triangular network – hitherto not paid much attention in the literature. A fascinating finding is that this compound exhibits characteristic features of PDA magnetism, also characterized by 1/3 magnetization plateau in this case, in a material with RKKY interaction, that too without any interference from other 4f-related phenomena (like the Kondo effect typical of Ce or U systems). It may be remarked that none of the isostructural ternary rare-earth compounds has been known to show such PDA characteristics in the past, in particular, other rare-earth compounds of this Rh family (R= Gd, Dy and Ho) (Ref. 68) as well as the Er counterpart in the Pd-based family, $Er_2PdSi_3$ [Ref. 56]. This is puzzling and renders support to the conclusion that this compound is a unique magnetic material, possibly suggesting an interesting magnetic interaction between orientated crystal-field-split ground state of 4f orbital of Er with Rh 4d-orbital and therefore it is worth probing this aspect further. Finally, this compound may also serve as a prototype for transport behavior across 1/3 magnetization plateau.

Authors acknowledge financial support from the Department of Atomic Energy (DAE), Govt. of India (Project Identification no. RTI4003, DAE OM no. 1303/2/2019/R&D-II/DAE/2079 dated 11.02.2020). E.V.S. thanks Department of Atomic Energy, Government of India, for awarding Raja Ramanna Fellowship. K.M. thanks financial support from BRNS, DAE under the DAE-SRC-OI program. S.M. thanks support from Vision Group on Science and Technology-GRD No 852.

Figure 1:

(a) Unit cell of $Er_2RhSi_3$ viewed along c-axis, showing triangular arrangement of Er ions in the Er layer; intralayer distances between Er atoms are given. An adjacent Rh-Si layer is also shown.
(b) X-ray diffraction pattern (Cu $K_\alpha$) below $2\theta= 40º$ to show weak superstructure lines (100) and (110). (c) Magnetic susceptibility as a function of temperature below 10 K for the zero-field cooled and field-cooled conditions of the specimens, measured with 10 Oe. Inverse $\chi$ in a field of 5 kOe is shown in the inset of (c), with a straight line through the Curie-Weiss regime.

Figure 2:
Real (top) and imaginary (bottom) parts of ac susceptibility of $Er_2RhSi_3$ below 6 K, obtained with different frequencies in zero-field as well as in 5 kOe. The curves for 5 kOe for all frequencies overlap and are featureless. Inset shows isothermal remnant magnetization at 2 K.

Figure 3:
(a) Zero-field and in-field heat-capacity below 10 K and (b) the zero-field heat-capacity in a wider temperature range of 2 - 40 K for $Er_2RhSi_3$. (b) The magnetic contribution to heat-capacity, $C_m$, for $Er_2RhSi_3$, employing the heat- capacity of $La_2RhSi_3$ (also shown) for non-magnetic part, along with the magnetic entropy ($S_m$) curve, derived from $C_m$ versus $T$ for the Er sample in the inset. (c) Isothermal entropy change curves as a function of temperature (see text) for different final fields starting from zero-field are plotted; the inset shows the electrical resistivity data at low temperatures.

Figure 4:
Isothermal magnetization of $Er_2RhSi_3$ below 60 kOe at 1.8, 4 and 6 K for the forward variation of the magnetic field. Inset shows the behavior in the range 60 – 120 kOe at 1.8 K. A schematic representation of a scenario (Ref. 5) for the orientation of magnetic moments in a triangular network; the question mark represents magnetically disordered ions.

Figure 5: Isothermal magnetoresistance, defined as $[\rho(H) - \rho(0)]/\rho(0)$, for $Er_2RhSi_3$, at (a) 2.2, (b) 4, (c) 10, (d) 15, and (e) 25 K.

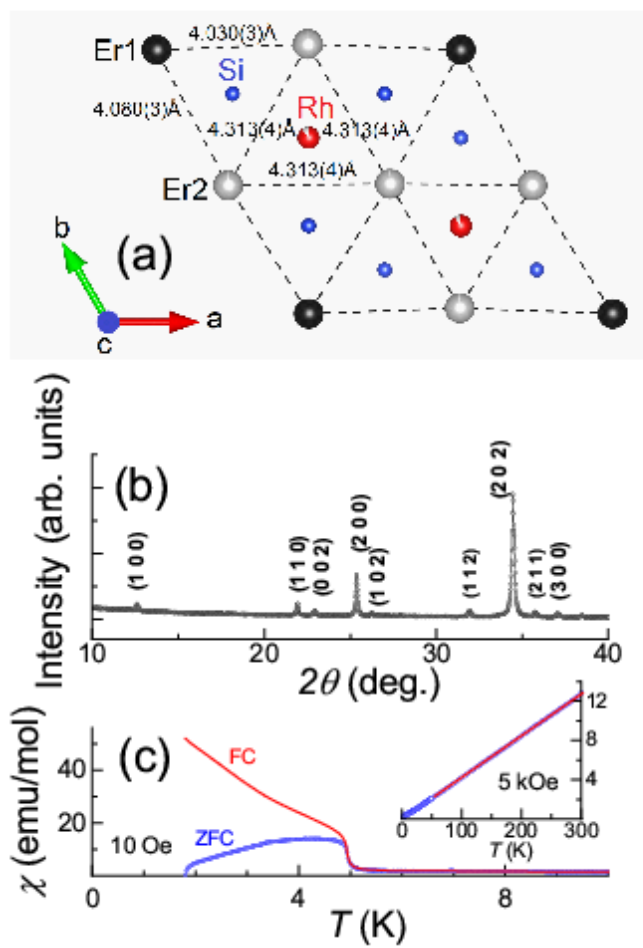

Figure 1

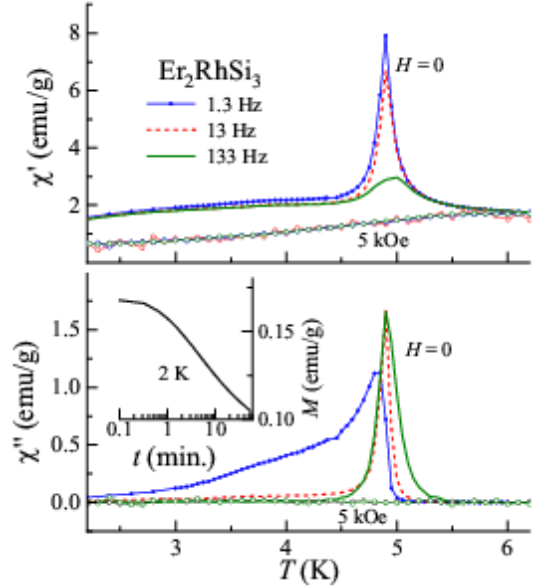

Figure 2

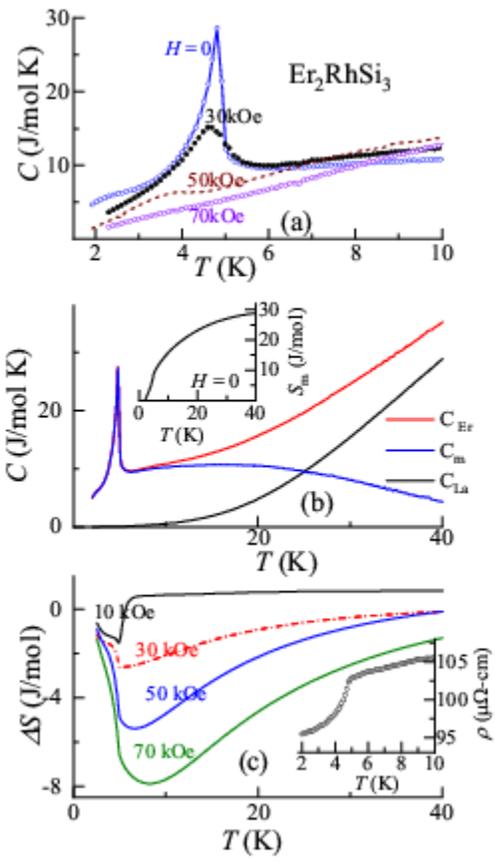

Figure 3

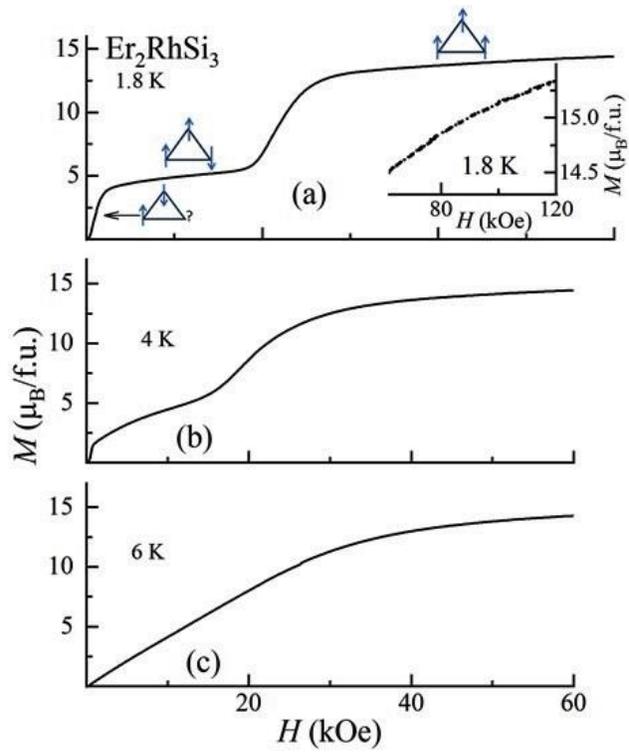

Figure 4

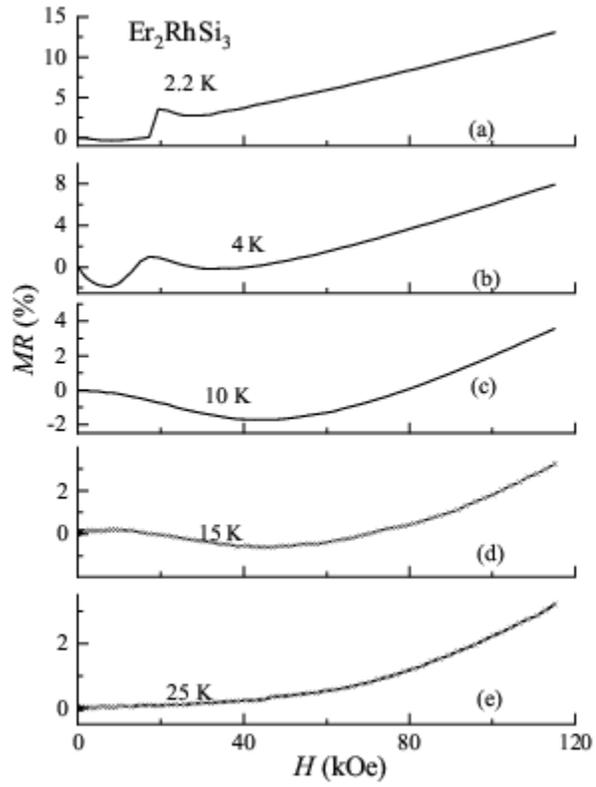

Figure 5

# Supplementary Material for
# Emergence of partially disordered antiferromagnetism and isothermal magnetization plateau due to geometrical frustration in a metallic compound, Er$_2$RhSi$_3$

Kartik K Iyer,[1,2] Kalobaran Maiti,[1] Sudhindra Rayaprol,[3] Ram Kumar,[4]
S. Mattepanavar,[5] S. Dodamani,[2] and E.V. Sampathkumaran[6]

[1]*Tata Institute of Fundamental Research, Homi Bhabha Road, Colaba, Mumbai – 400005, India*
[2]*KLE Society's Dr. Prabhakar Kore Basic Science Research Centre, KLE Academy of Higher Education and Research, Belagavi- 590010, India*
[3]*UGC-DAE-Consortium for Scientific Research -Mumbai Centre, BARC Campus, Trombay, Mumbai – 400085, India*
[4]*Maryland Quantum Materials Center, University of Maryland, College Park, MD20742, USA.*

[5]*KLE Society's, Basavaprabhu Kore Arts, Science & Commerce College Chikodi-591201, India.*
[6]*Homi Bhabha Centre for Science Education, TIFR, V. N. Purav Marg, Mankhurd, Mumbai – 400088, India.*

Here we present the experimental methods details, crystallographic features of Er$_2$RhSi$_3$, as reported in Ref. S1, and Rietveld fitting for our x-ray diffraction patterns. Magnetic susceptibility behavior around the magnetic transition measured with a few low fields for the zero-field-cooled (ZFC) and field-cooled (FC) conditions is also presented.

Dc susceptibility ($\chi$) measurements in various dc magnetic fields were performed as a function of temperature (*T*) with the help of a superconducting quantum interference device (Quantum Design); the same magnetometer was used to measure ac $\chi$ with various frequencies. Isothermal magnetization curves were recorded with a commercial vibrating sample magnetometer (Quantum Design) at different temperatures in the magnetically ordered state. Heat-capacity and resistivity measurements were done using a (Quantum Design) Physical Properties Measurements System in zero field as well as in the presence of different dc magnetic fields. All the measurements were performed while warming from 1.8 K for the zero-field-cooled condition of the specimen (unless otherwise stated).

Views of crystal structure in different orientations are shown in Fig. S1. This compound crystallizes in a AlB$_2$-derived hexagonal structure. It is generally difficult to decide whether the space group is *P$\bar{6}$2c* or *P*6$_3$/*mmc* [S1, S2]. It is clear from the Fig. S1 that Rh and Si form a honeycomb network and these layers are intercalated by Er layers. Due to the crystallographic ordering of Si and Rh, one type of Er (called Er1) is surrounded by 12 Si atoms (6 each from hexagons nearby honeycomb layers), and the other type Er arises (called Er2) due to the surrounding by two hexagons each containing 4Si and 2Rh ions. Possibly due to this different chemical surrounding of the Er ions in this manner, x and y coordinates for Er2 are reported to get distorted (as shown in Fig. S2b), based on crystallographic analysis of single crystals at room temperature [S1]. Also see Table S1 reproduced from Ref. S1. This results in subtle variations in the interatomic distances, the details of which can be found in Ref. S1. It is not obvious to us at present whether such subtle variations are responsible for some kind of PDA magnetic structure. Therefore, it would be of interest to focus studies on crystallographic analysis in the magnetically ordered state. It is clear from Fig. S3c that Er ions form a triangular network.

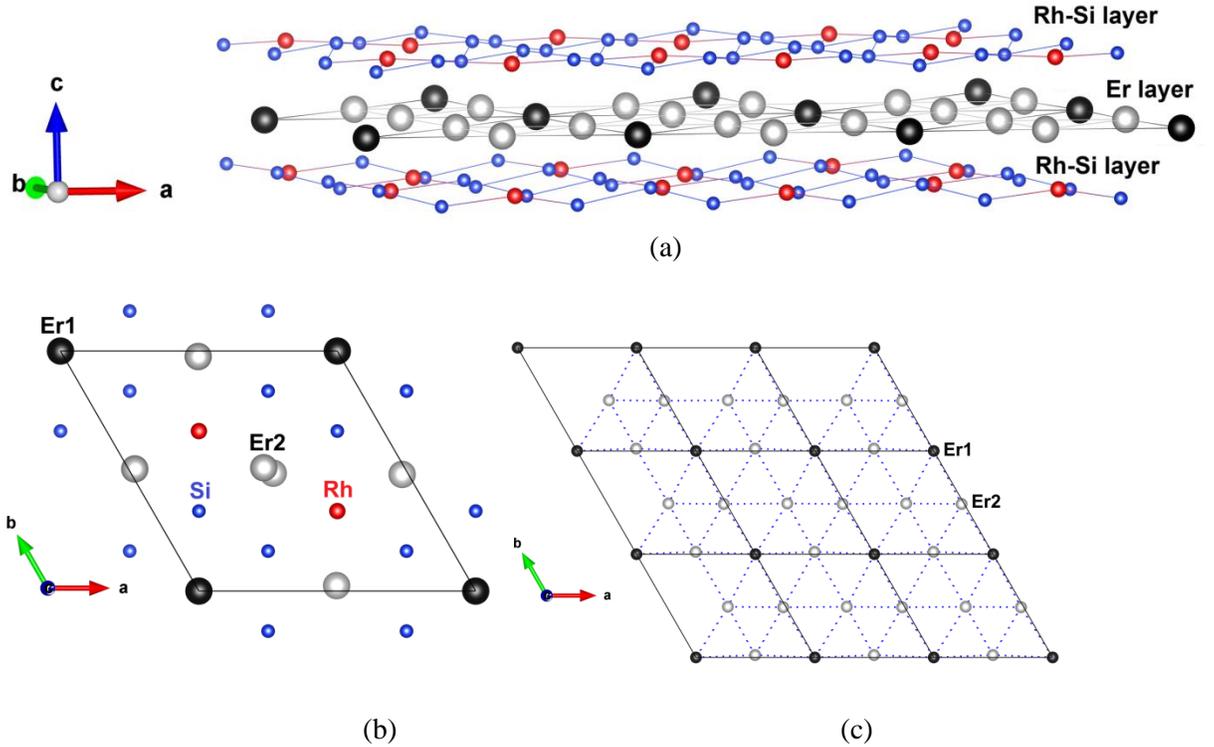

(a)

(b)                                               (c)

Fig. S2. Crystallographic views of $Er_2RhSi_3$ along (a) almost b-direction to show the honeycomb layers of Rh-Si, intercalated by Er, (b) c-direction showing unit-cell with two layers of Er with an intervening Rh-Si layer to bring out a small deviation in the coordination of Er2, and (c) c-direction showing one Er layer to bring out triangular network of Er ions within a layer.

**Table S1: Atomic coordinates of the atoms in the $Er_2RhSi_3$ lattice (reproduced from Ref. S1)**

| Atom  | Position | x       | y       | z    |
|-------|----------|---------|---------|------|
| Er (1)| 2(b)     | 0.00000 | 0.00000 | 0.25 |
| Er (2)| 6(h)     | 0.481   | 0.019   | 0.25 |
| Rh    | 4(f)     | 0.33333 | 0.66667 | 0    |
| Si    | 12(i)    | 0.167   | 0.3333  | 0    |

Fig.S2 shows the x-ray diffraction pattern in a wider angle range. The pattern was subjected to Rietveld analysis and the parameters obtained are also included in this figure, along with indexation of the peaks. It is distinctly clear that there are superstructure lines (100) and (110), establishing doubling of unit-cell parameters with respect to the unit-cell for $AlB_2$-like crystal structure.

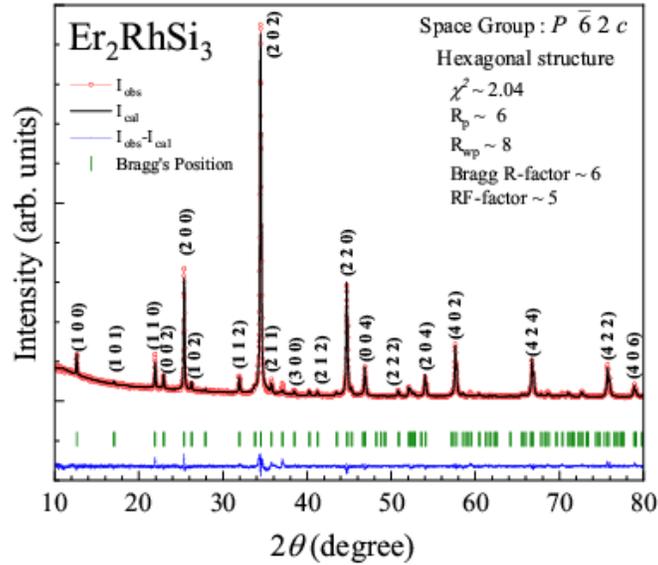

Fig. S2: X-ray diffraction pattern of $Er_2RhSi_3$ (Cu $K_\alpha$) in the range $2\theta$= 10-80° with Rietveld fitting and the fitted parameters

In Figure S3, we show the magnetic susceptibility behavior measured with ($H=$ )10, 20, 50 and 100 Oe in the range 2-20 K, to show the feature around magnetic transition. A notable finding is that the temperature at which these curves deviate reduces with increasing $H$; for instance, with 100 Oe measuring field, the deviation occurs at about 3.2 K, while at 10 Oe, the curves diverge at the onset of magnetic order itself, as though possible spin-glass freezing occurs at 5 K in the absence of any external field. Clearly, there is a high sensitivity of this characteristic freezing temperature to even small applications of dc $H$ (that is, undergoing a sharp downward shift).

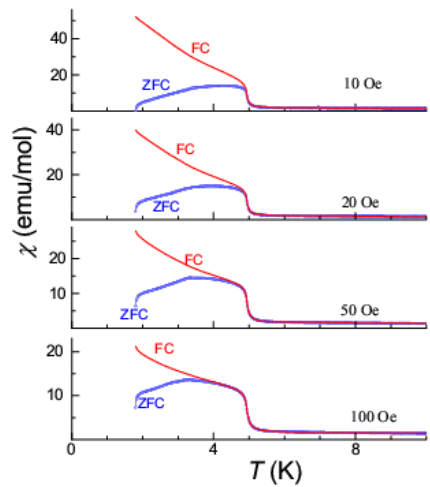

Fig. S3: Magnetic susceptibility as a function of temperature below 20 K measured with 10, 20, 50 and 100 Oe for $Er_2RhSi_3$.